\documentclass[aps,prl,reprint,groupedaddress]{revtex4-2}
\usepackage{graphicx}
\usepackage{siunitx}
\usepackage{hyperref}
\usepackage{epstopdf}

\begin{document}


\title{Gate induced strain on a two-dimensional hole gas in silicon}


\author{D. van der Bovenkamp}
\affiliation{MESA+ Institute for Nanotechnology, University of Twente, P.O. Box 217, 7500 AE Enschede, The Netherlands}
\email{d.vanderbovenkamp@utwente.nl}
\author{C.S.A. Müller}
\affiliation{MESA+ Institute for Nanotechnology, University of Twente, P.O. Box 217, 7500 AE Enschede, The Netherlands}
\author{B.D. Pantiru}
\affiliation{MESA+ Institute for Nanotechnology, University of Twente, P.O. Box 217, 7500 AE Enschede, The Netherlands}

\author{I. Bo\v{s}njak}
\affiliation{MESA+ Institute for Nanotechnology, University of Twente, P.O. Box 217, 7500 AE Enschede, The Netherlands}

\author{M. Cignoni}
\affiliation{MESA+ Institute for Nanotechnology, University of Twente, P.O. Box 217, 7500 AE Enschede, The Netherlands}

\author{Q. Torrent Nicolau}
\affiliation{MESA+ Institute for Nanotechnology, University of Twente, P.O. Box 217, 7500 AE Enschede, The Netherlands}

\author{M.E. Bal}
\affiliation{High Field Magnet Laboratory (HFML-EMFL), Radboud University, Toernooiveld 7, 6526 ED Nijmegen, The Netherlands}
\affiliation{Radboud University, Institute for Molecules and Materials, 6525 AJ Nijmegen, The Netherlands}

\author{S. Wiedmann}
\affiliation{High Field Magnet Laboratory (HFML-EMFL), Radboud University, Toernooiveld 7, 6526 ED Nijmegen, The Netherlands}
\affiliation{Radboud University, Institute for Molecules and Materials, 6525 AJ Nijmegen, The Netherlands}

\author{J. Ridderbos}
\affiliation{MESA+ Institute for Nanotechnology, University of Twente, P.O. Box 217, 7500 AE Enschede, The Netherlands}
\author{F.A. Zwanenburg}
\affiliation{MESA+ Institute for Nanotechnology, University of Twente, P.O. Box 217, 7500 AE Enschede, The Netherlands}


\date{\today}

\begin{abstract}
We show the effect of gate-induced strain on the valence band of a silicon (Si) metal oxide semiconductor (MOS) confined two-dimensional hole gas (2DHG). Increasing aluminum gate thickness, and thereby the strain in the channel, results in the onset of a second subband contributing to Shubnikov-de Haas oscillations. Temperature-dependent magnetotransport measurements reveal distinct cyclotron masses of $m_c^*=(0.36\pm0.04)m_0$ and $m_c^*=(0.49\pm0.02)m_0$. The measured cyclotron masses differ from those expected for an idealized heavy-hole (HH)/light-hole (LH) picture, reflecting the combined influence of quantum confinement, strain, and HH–LH mixing on the valence band.
\end{abstract}


\maketitle

\section{}
Hole spin qubits in silicon (Si) have gained more interest following the success  of holes in germanium (Ge) and Ge/SiGe heterostructures \cite{Scappucci2020, Zwanenburg2013}. Their p-type symmetry suppresses hyperfine interaction, making them less susceptible to environmental noise of nuclear spins \cite{Bulaev2007}. The strong spin orbit interaction (SOI) of holes allows electric driving, making micromagnets redundant \cite{Piot2022}. High-fidelity single and two-qubit gates in Ge have been established as well as multi-qubit arrays showing fast Rabi oscillations \cite{Wang2022,Wang2024,John2025, Hendrickx2020, Hendrickx2021}. 
Both Si and Ge have a fourfold degeneracy at the top of the valence band which needs to be lifted to create a two-level quantum system. For Ge, this fourfold degeneracy at the $\Gamma$ point can be lifted by strain controlled lattice mismatch in GeSi buffer layers to realize an almost pure heavy hole (HH) ground state \cite{Terrazos2021}. In Si the degeneracy is lifted by confinement, however a pure HH ground state is hard to achieve due to the mixing between HHs and light holes (LH) \cite{Donetti2011}. The downside of a well-isolated HH state is a highly anisotropic g-factor which makes large qubit arrays susceptible to the orientation of the magnetic field. The mixing of HHs and LHs leads to less anisotropy which can improve the overall driving speed of larger structures \cite{Mauro2025}.

Recent advancements focus on strain tunability in germanium to increase SOI and potentially HH-LH mixing \cite{Sarkar2025, Dsouza2024, Vecchio2024}. Si has an already mixed state due to confinement and recently the effect of mixing to increase SOI has been investigated \cite{Stano2025}. Strain is a powerful tool for increasing mixing and altering the valence-band structure, but it is harder to apply in Si, since Si devices typically do not rely on heterostructures. Therefore, the focus for Si is to use gate-induced strain \cite{Liles2021}. Using gataes to create strain results in an inhomogeneous strain profile which opens the possibility for driving g-tensor modulation resonance \cite{Abadillo2023}. Gate-induced strain can also be used to induce HH-LH splitting in systems like acceptor-based qubits where the confinement does not break crystal symmetry \cite{VanderHeijden2014, Mol2015}. In such systems it is even possible to use strain to tune between a LH \cite{Salfi2015} and HH \cite{Salfi2016} ground state. Fine tuning of strain allows operation at sweet spots, making the qubit states less sensitive to electric field noise \cite{Zhang2023}. 

In this letter, we show how local gate-induced strain alters the valence-band structure of a silicon two-dimensional hole gas. We use the mismatch in thermal expansion coefficient between silicon and aluminum (Al) to create local strain in the silicon lattice at cryogenic temperatures and vary the thickness to control the magnitude of the strain. Because the strain profile is defined by a local Al gate, this approach enables local band-structure engineering using a standard fabrication process. In addition, it is also a straightforward method to implement local strain controlled by the thickness of the Al gate. We probe the effect of strain on the valence band by extracting the Shubnikov-de Haas frequencies and effective cyclotron masses, which reflect the subband occupation and band curvature set by strain-induced HH-LH mixing.

Hall bars are fabricated on intrinsic silicon with a silicon dioxide (SiO$_2$) dielectric of \SI{100}{nm}. The width of the Hall bar is \SI{10}{\micro m} with a distance between longitudinal contacts of \SI{80}{\micro m}. The source and drain, as well as the hall contacts, are p-type (boron) doped with a dose of $5\times10^{14}$ cm$^{-2}$. Structures are patterned using optical lithography to prevent mobility reduction from E-beam lithography \cite{Kim2017}. E-beam evaporation is used to deposit Al gates of varying thickness, which is followed by lift off in hot dimethyl sulfoxide (DMSO). The final step is a high-temperature steam anneal (HSA) at \SI{400}{\degree C} for 5 minutes to passivate defects in the Si/SiO$_2$ interface \cite{Abe2001}.

Measurements are performed at a temperature of \SI{1.3}{K} and a magnetic field perpendicular to the device. A low-frequency ac voltage is applied via high-ohmic pre-resistor to realize a current-driven source-drain bias. The resistance data were acquired in four-probe configuration with a constant current excitation of \SI{10}{nA} using standard lock-in acquisition. The gate is controlled via a dc voltage, which accumulates a two-dimensional hole gas (2DHG) at the Si/SiO$_2$ interface, and controls the carrier concentration in the channel.

\begin{figure}
    \includegraphics[width=0.98\linewidth]{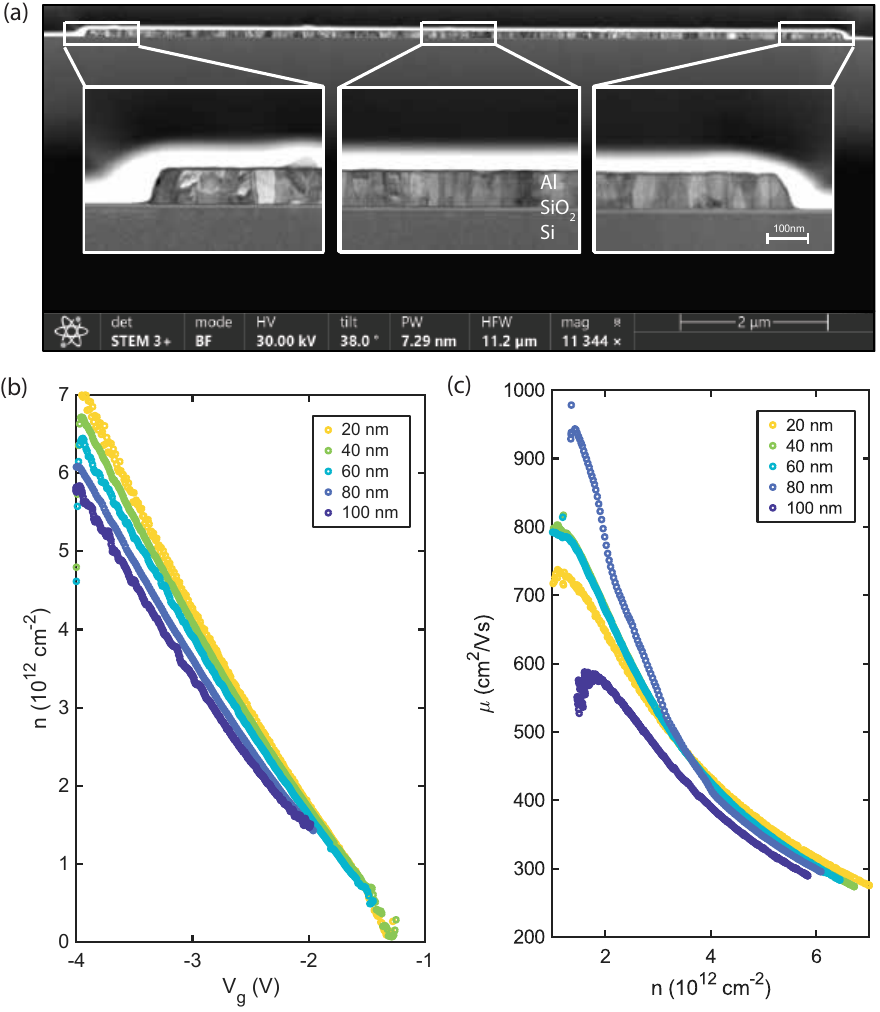}
    \caption{(a) cross-sectional scanning transmission electron microscopy image of the aluminum metal gate in the channel direction with insets of the sides and center. (b) Hall carrier density $n_H$ as function of gate voltage $V_g$ for varying gate thicknesses from 20-\SI{100}{nm}. (c) Mobility $\mu$ as a function of carrier density for varying gate thicknesses from 20-\SI{100}{nm}. Both measurements are taken at a magnetic field of \SI{9}{T} perpendicular to the device. The device with a gate thickness of \SI{100}{nm} is measured at $T=$ \SI{1.3}{K} while the other thicknesses are measured at $T=$ \SI{1.7}{K}.}
    \label{fig:ns_mu}
\end{figure}

A cross-sectional scanning transmission electron microscopy (STEM) image of the Al gate is shown in Fig. \ref{fig:ns_mu}(a), together with magnified views of the gate center and edges. The cross-section is perpendicular to the transport direction, showing a \SI{10}{\micro\meter}-wide gate structure. The Al gate exhibits a polycrystalline structure with grain sizes on the order of several tens of nanometers. Despite this local variation, the gate is expected to generate on average a smooth compressive strain profile underneath the gate where shear strain is dominant at the edges. 

The device is characterized by Hall carrier density $n_H$ and hole mobility $\mu$. The carrier density versus gate voltage $V_g$ is shown in Fig. \ref{fig:ns_mu}(b) for devices with a gate thickness of 20, 40, 60, 80 and \SI{100}{nm}. The device with a gate thickness of \SI{100}{nm} is measured at $T=$ \SI{1.3}{K} and the other thicknesses are measured at $T=$ \SI{1.7}{K}. We expect this small deviation in temperature to be negligible and not affect the measurements significantly. A linear fit results of an average oxide capacitance of $3.8\times10^{-3}$ F/m$^2$ and an average oxide thickness $t_{ox} \approx$ \SI{9.2}{nm}, in agreement with ellipsometry measurements and the STEM image. The mobility as function of carrier density is shown in Fig. \ref{fig:ns_mu}(c), where the peak mobility varies between \SI{600}{cm^2/Vs} and \SI{940}{cm^2/Vs} for different gate thicknesses. Because we observe variation also across different devices with the same gate thickness, this device variation is most likely due to difference in surface-roughness scattering at the Si/SiO$_2$ interface \cite{Sabbagh2019}. The hole mobility is relatively low compared to typical electron mobilities due to the nonparabolicity of the valence band \cite{Fischetti2003}, which results in a larger transverse effective mass \cite{Wendoloski2026}. Therefore, high magnetic fields are required to observe SdH oscillations.

\begin{figure}
    \centering
    \includegraphics[width=\linewidth]{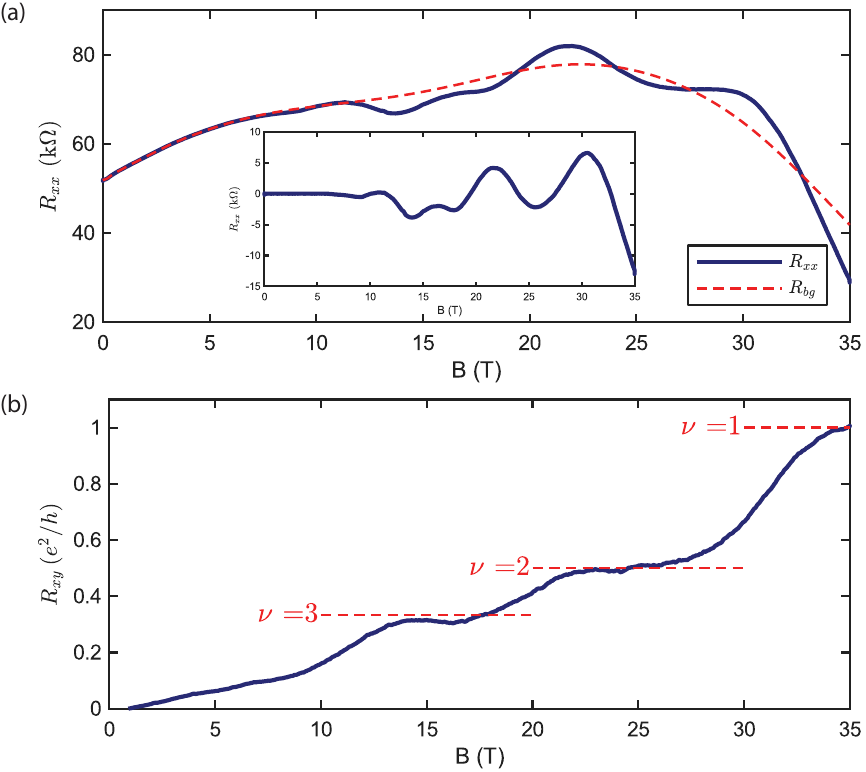}
    \caption{(a) Longitudinal resistance $R_{xx}$ versus magnetic field. The background magneto resistance fit is shown in the red dashed line. The background removed resistance is shown in the inset. (b) Hall resistance $R_{xy}$ in units of $e^2/h$ as function of magnetic field. Red dashed lines are a guide to the eye to indicate the expected quantized values corresponding to filling factors $\nu=1,2,3$. Both measurements are obtained at $T=$ \SI{1.3}{K} with $V_g = $ \SI{-1.95}{V} and for a gate thickness of \SI{100}{nm}.}
    \label{fig:rxx_rxy}
\end{figure}

We study magnetic field dependence of the longitudinal resistance $R_{xx}$, as shown in Fig. \ref{fig:rxx_rxy}(a). The spline fitted background magnetoresistance $R_{bg}$, shown as red dashed line, is subtracted from the measured magnetoresistance and the background-removed signal is shown in the inset. 
Due to confinement and applied strain, we expect two types of charge carriers with different concentrations and mobilities. Consequently, we can no longer fit the background resistance with a polynomial and should use a two-carrier model \cite{Dill2025}. However, the resistance drop when reaching the quantum limit around \SI{26}{T} prevents the background resistance to be described by the two-carrier model from Ref.\cite{Dill2025}, hence the use for spline fit. The approach of the quantum limit can also be observed in Fig. \ref{fig:rxx_rxy}(b) where the Hall resistance $R_{xy}$ in units of $e^2/h$ is shown versus magnetic field. The red dashed lines represent the quantized plateaus \cite{Klitzing1980}:
\begin{eqnarray}
    R_{xy} = \frac{h}{\nu e^2},
\end{eqnarray}
where $e$ is the elementary charge and $h$ is the Planck constant. The filling factor $\nu = 1,2,3$ is used to emphasize the integer quantum hall plateaus. At high magnetic field the quantum limit is approached as the filling factor 1 is achieved. The observed odd filling factors indicates that the Landau levels are sufficiently resolved such that spin degeneracy is lifted at high magnetic fields.

\begin{figure}
    \centering
    \includegraphics[width=0.98\linewidth]{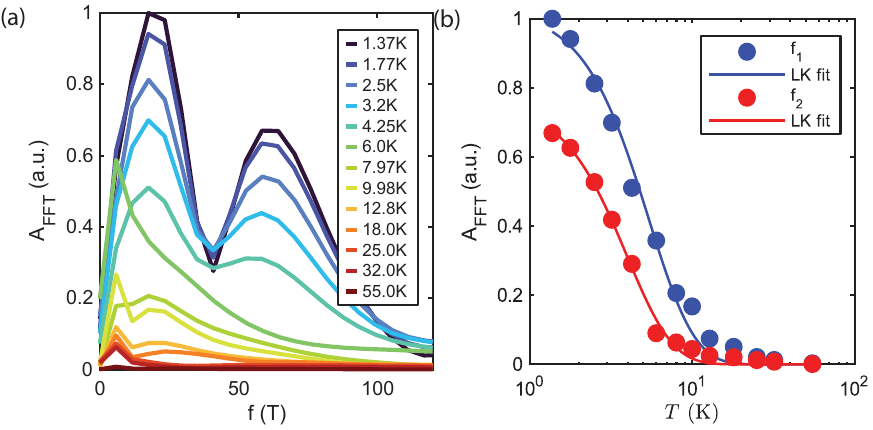}
    \caption{(a) Temperature dependent discrete fast Fourier amplitude $A_{FFT}$ of the SdH oscillations versus frequency for a gate thickness of \SI{100}{nm}. (b) FFT amplitude as a function of temperature for $f_1 \approx 19$ T and $f_2\approx 66$ T. The solid lines represent the fit of the Lifshitz-Kosevich thermal dampening factor. }
    \label{fig:fft_meff}
\end{figure}

To create the FFT spectrum in Fig.\ref{fig:fft_meff}, we remove the background magnetoresistance from the longitudinal resistance and use a Savitzky–Golay smoothing filter. We observe two peaks at $f_1 \approx 19$ T and $f_2 \approx 66$ T for temperatures from 1.3-55 K at a $V_g = -2.1$ V. To determine the effective cyclotron mass corresponding to $f_1$ and $f_2$, we analyze the temperature dependence of the SdH oscillation amplitude $A_{FFT}$. The amplitude of the peaks at $f_1$ and $f_2$ are plotted versus temperature in Fig. \ref{fig:fft_meff}(b) where the amplitude reduction is fitted according to the Lifshitz-Kosevich(LK) thermal dampening factor $A(T)$ as \cite{Lifshitz1956}:

\begin{eqnarray}
    A(T) = \frac{\alpha T m^* /B_{FFT}}{\sinh(\alpha T m^* /B_{FFT})},
\end{eqnarray}

where  $\alpha= \frac{2\pi^2 k_B}{e\hbar}$ is the LK prefactor. $T$ denotes the temperature, $\hbar$ the reduced Planck constant, and $k_B$ the Boltzmann constant. $B_{FFT}$ corresponds to the logarithmic average magnetic field over the FFT window (5-35 T). The effective cyclotron masses obtained from the LK fit are $m_c^* \approx 0.36 \pm 0.04$ for $f_1$, and $m_c^* \approx 0.49 \pm 0.02$ for $f_2$. This disparity in effective cyclotron masses is in agreement with the presence of two distinct subbands. In anisotropic band structures, the cyclotron mass represents an orbit-averaged quantity and may differ from direction-dependent band masses or density-of-states effective masses reported in the literature \cite{Donetti2011}. Rather than corresponding directly to bulk LH and HH masses, these values reflect subband-dependent in-plane masses of a mixed valence band \cite{Stano2025}. In confined silicon 2D hole systems, HH-LH mixing due to shear strain and quantum confinement can significantly renormalize the dispersion \cite{Jianli2011, Sun2007, GSun2007}, even to the extent where the lower subband exhibits a HH-like mass, while the higher subband retain a lighter, more LH-like mass. Therefore, we refer to them as the first and second subbands rather than assigning LH and HH character.

\begin{figure}
    \centering
    \includegraphics[width=0.98\linewidth]{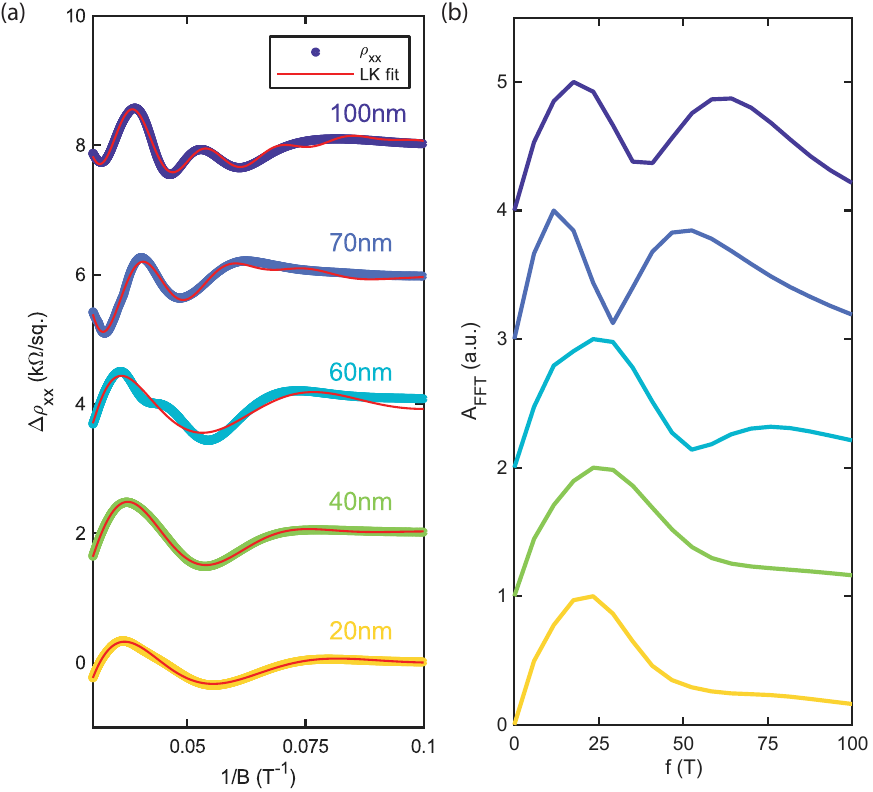}
    \caption{(a) Normalized longitudinal resistivity after background removal and smoothing as function of inverse magnetic field $1/B$. The baseline subtracted resistivity $\Delta\rho_{xx}$  is shown for varying gate thicknesses from 20-100 nm and offset to improve visibility. Measurements are performed at $T=1.3$ K and a gate voltage of \SI{-2.1}{V} corresponding to $n_H \approx 1.7\times10^{12}$. The red line presents the Lifshitz-Kosevich fit.
    (b) Normalized discrete fast Fourier transform amplitude $A_{FFT}$ as a function of frequency $f$. Colors denote the thickness of the metal gate corresponding to the color shown in (a).}
    \label{fig:thickness_comparison}
\end{figure} 

The effect of gate thickness on the SdH oscillations is shown in Fig. \ref{fig:thickness_comparison}. The longitudinal resistance is converted to resistivity $\rho_{xx}$ where $\Delta\rho_{xx} = \rho_{xx} - \rho_0 $, with $\rho_0$ being the low field resistivity, and plotted versus $1/B$ to highlight periodicity of the SdH oscillations. The oscillation amplitude follows an exponential envelope indicating a singular conduction channel in the 2DHG \cite{Lodari2022}. Increasing the Al gate thickness results in the appearance of a beating pattern in the SdH oscillations where, the gate voltage is kept at \SI{-2.1}{V} to set the carrier density similar across all thicknesses and not influence confinement splitting. 
We fit the oscillatory component in $1/B$ using the LK expression \cite{Lifshitz1956}:

\begin{eqnarray}
\frac{\Delta R_{xx}}{R_0} =
\sum_{i=1}^{f}
A_i
R_T
R_D^{(i)}
R_S^{(i)}\cos\!\left[2\pi\left(\frac{F_i}{B}-\phi_i\right)\right].
\end{eqnarray}

\noindent
The frequency is proportional to the extremal Fermi surface $A_F$ via $F_i = \frac{\hbar}{2\pi e}A_F^{(i)}$ and $\phi_i$ is the phase. We use thermal dampening $R_T$, dingle dampening $R_D$ and spin splitting $R_S$ which are elaborated in Appendix A\ref{sec:appA}. The effective mass is fixed using the obtained values from the thermal damping on FFT amplitude. The quantum oscillations from Fig. \ref{fig:thickness_comparison}(a) are fitted per thickness, shown in red. For gate thicknesses below 60 nm the oscillatory component $\Delta\rho_{xx}$ is well described by a single-frequency LK term. For thicker gates, a satisfactory fit requires the inclusion of two frequency components. The fitted frequencies are \SI{15.5}{T} and \SI{65.6}{T} for a gate thickness of \SI{100}{nm} which is in good agreement with the FFT signal in Fig. \ref{fig:thickness_comparison}(b). A large spread is observed for the dingle temperature ($T_D \approx 3-10$ K) and g-factor ($g^*\approx3-6$) likely due to the many fitting parameters in combination with the low number of oscillations. Increasing strain also affects the g-factor significantly and alters the valence band structure \cite{Liles2021}. Within the Bir-Pikus formalism \cite{Pikus1974}, biaxial strain lifts the HH-LH degeneracy, while shear strain introduces off-diagonal coupling terms that induce HH/LH hybridization. Finite-element simulations of the strain field generated by the Al gate, presented in Appendix B, show the presence of biaxial strain underneath the Al gate and shear strain components towards the edges. The simulated strain magnitudes are sufficient to modify the HH/LH energy landscape on the meV scale, suggesting that the observed evolution of the SdH spectrum is consistent with a strain-induced reconstruction of the valence-band structure. These changes reshape and anisotropically renormalize the g-tensor\cite{Liles2021, Crippa2018}. The spin splitting term in the LK fit depends on the product $g^*m^*$, so any variation in the fitted $g^*$ may also be related to the contributions from the difference in effective cyclotron masses.
The effect of the occupation of a second subband is more clarified in the corresponding FFT spectra shown in Fig. \ref{fig:thickness_comparison}(b). For an Al gate thickness of \SI{20}{nm}, a singular peak at $f_1 \approx 19$ T indicates a single occupied subband. A second FFT peak starts to arise at $f_2 \approx 66$ T for an Al gate thickness of \SI{60}{nm} which becomes more pronounced for increasing thickness. The second peak is not located at a harmonic frequency of $f_1$ and the separation between the two peaks is too large to be attributed to spin splitting. The emergence of a second frequency occurs systematically with increasing gate thickness despite nearly identical Hall carrier densities.

\begin{figure}
    \centering
    \includegraphics[width=0.98\linewidth]{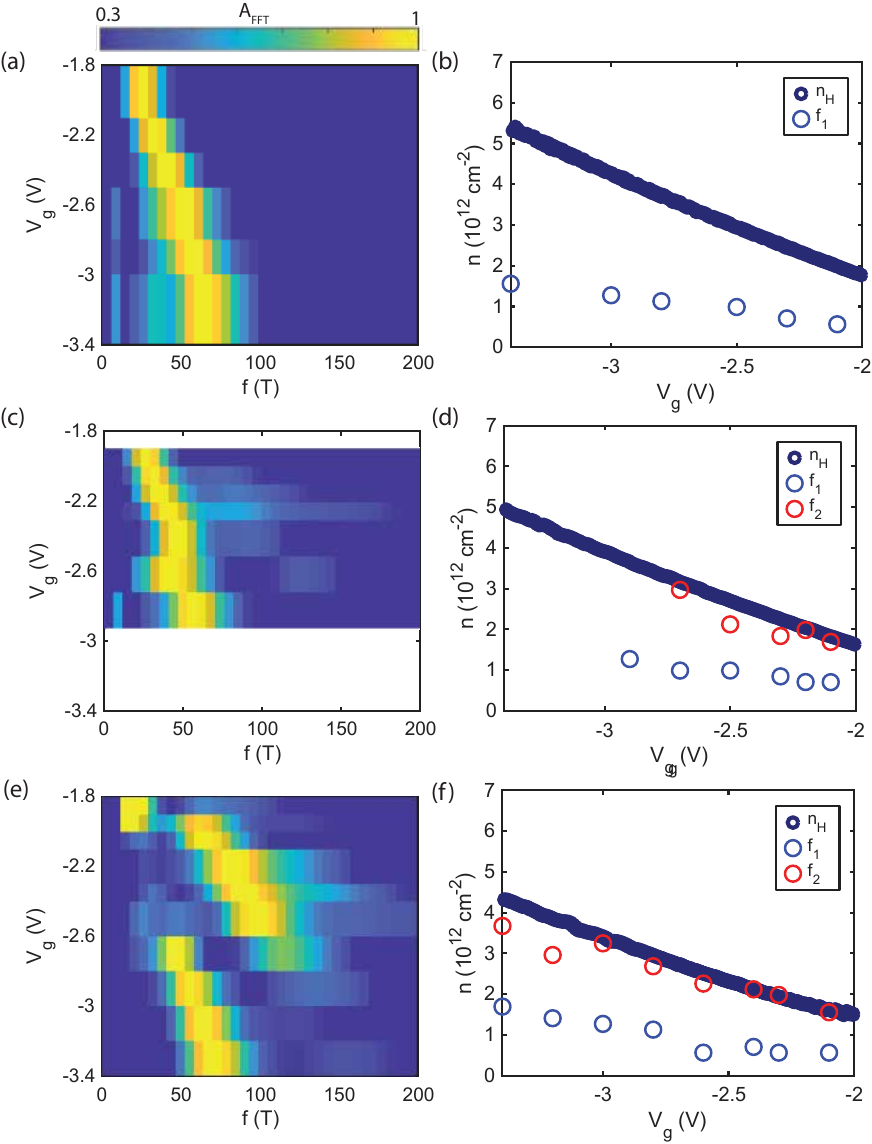}

    \caption{FFT amplitude as function of frequency $f$ and gate voltage $V_g$ for an Al gate thickness of (a) \SI{20}{nm}, (c) \SI{60}{nm} and (e) \SI{100}{nm} at $T=1.3$K. The amplitude color scale is set from 0.3 to 1. The corresponding carrier densities obtained from the FFT peak positions are shown as circles in (b), (d), and (f), together with the Hall density $n_H$ obtained from Fig.\ref{fig:ns_mu}(b).}
    \label{fig:fft_vg}
\end{figure}

The gate-voltage dependence of the FFT spectrum is shown in Fig. \ref{fig:fft_vg} for devices with Al gate thicknesses of (a) \SI{20}{nm}, (b) \SI{60}{nm}, and (c) \SI{100}{nm}. The FFT frequencies are converted to carrier densities using $n=g_zg_v(e/h)f$, where $g_z=1$ and $g_v=1$ denote the spin and valley degeneracies, respectively. Here, we assume $g_z=1$ as the spin degeneracy is lifted in the magnetic field range under consideration (see Fig. \ref{fig:rxx_rxy}(b)). 

Interestingly, the carrier density associated with $f_2$ closely follows the Hall density n$_H$ obtained from Fig.\ref{fig:ns_mu}(b), whereas $f_1$ accounts for only a smaller fraction of the total carrier density. Assuming we have two carrier transport with different mobility, the single-carrier Hall density is a mobility-weighted average, rather than the sum $n_1 + n_2$ \cite{Dill2025}. Therefore, the density is biased toward whichever channel dominates conductivity. As gate thickness and strain increase, the second subband's contribution to conductivity grows. This results in $n_H$ approaching $n_{f_2}$. Although $f_1$ is resolved by SdH oscillations for devices with a thinner gate, the discrepancy with $n_H$ indicates that a second channel already contributes to transport for thinner gates. However, the mobility of this channel is most likely too low to be resolved in SdH oscillations.

In conclusion, we show the tunability of the valence band in a silicon two-dimensional hole gas using local strain induced by an aluminum metal gate. Increasing the gate thickness induces a clear beating pattern in the SdH oscillations together with the onset of a second FFT peak, indicating the occupation of two subbands. The effective cyclotron masses extracted from thermal FFT amplitude dampening are distinctive for each subband. The obtained values differ from bulk values for HHs LHs and represent a mixed valence band state. These observations are consistent with strain- and confinement-induced modifications, where HH-LH mixing renormalizes the effective masses and subband energies. Our results demonstrate that the aluminum gate thickness provides an effective tuning parameter for engineering the subband occupation in a Si 2DHG which can be usefull for hole spin qubit applications.

\section{Appendix A:}
\label{sec:appA}
The thermal dampening dingle dampening and spin splitting are described by:

\begin{eqnarray}
&&R_T= \frac{X_i}{\sinh X_i}, \qquad
R_D^{(i)} = \exp\!\left(-\frac{2\pi^2k_BT_D^{(i)} m^*_{(i)}}{\hbar e B}\right), \nonumber \\
&&R_S^{(i)} = \cos\left(\frac{\pi g^*_{(i)} m^*_{(i)}}{2m_e}\right), \nonumber
\end{eqnarray}

where $k_B$ is the Boltzmann constant, $\hbar$ is the reduced Planck constant and $m_e$ the electron rest mass. For holes the effective g-factor is generally anisotropic and described by a tensor. For this fit, we use an effective scalar $g^*$ as the projection of the g-tensor along the magnetic field direction. The thermal dampening factor and dingle temperature are given by:

\begin{eqnarray}
X_i = \frac{2 \pi^2 k_B  T}{\hbar\omega_c^{(i)}},
\qquad
T_D^{(i)} = \frac{\hbar}{2 \pi k_B \tau_{(i)} }.\nonumber
\end{eqnarray}
Here $T$ is the temperature and $T_D$ is the dingle temperature, $\omega_c$ the cyclotron frequency, and $\tau$ the scattering time.

\section{Appendix B:}
\label{sec:appB}
\begin{figure}
    \centering
    \includegraphics[width=0.98\linewidth]{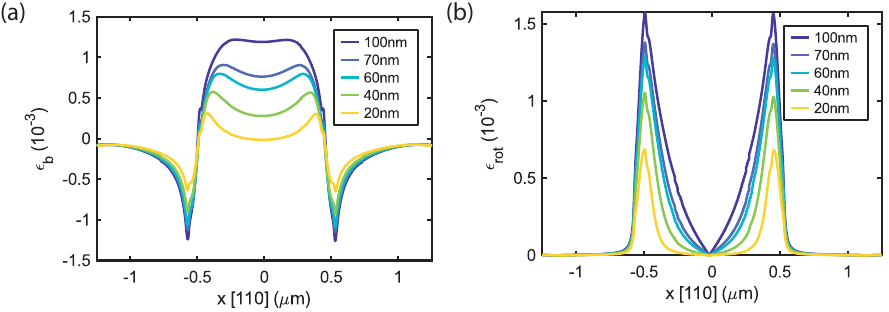}
    \caption{(a) biaxial strain invariant and (b) shear strain component obtained from COMSOL Multiphysics strain tensor for varying gate thicknesses from 20 to 100 nm.}
    \label{fig:Simulations}
\end{figure}

Device simulations have been performed using the COMSOL Multiphysics finite-element solver. The simulated layer stack is constructed based on the STEM cross section shown in Fig. \ref{fig:ns_mu}a. Thermal contraction was simulated by cooling the structure from room temperature to $T = $ \SI{40}{mK}, assuming a strain-free state at room temperature. The strain tensor components were extracted \SI{1}{nm} below the Si/SiO$_2$ interface along the full width of the gate. To reduce computational time, a \SI{1}{\micro m}-wide gate is simulated instead of the experimentally realized \SI{10}{\micro m} gate. Since the strain field varies primarily near the gate edges, this reduction does not affect the qualitative conclusions of the analysis.
The strain distribution obtained from simulations is analyzed in terms of the biaxial strain invariant shown in Fig. \ref{fig:Simulations}a

\begin{eqnarray}
\epsilon_b=\epsilon_{zz}-\frac{\epsilon_{xx}+\epsilon_{yy}}{2},
\end{eqnarray}

which constitutes the symmetry-breaking strain component entering the Bir-Pikus Hamiltonian for the silicon valence band \cite{Pikus1974}. This term lifts the HH/LH degeneracy and produces an estimated splitting

\begin{eqnarray}
\Delta_{HL}\approx 2b\epsilon_b,
\end{eqnarray}

where $b=-2.1~\mathrm{eV}$ is the valence-band deformation potential of silicon \cite{Sun2007}. The simulated strain field reaches $\epsilon_b\approx1.2\times10^{-3}$ beneath the Al gate, corresponding to an estimated HH-LH splitting of approximately \SI{5}{meV}. This estimate should be regarded as an order-of-magnitude value only. The relation describes the strain-induced splitting of the bulk valence bands at the zone center, whereas the present system consists of a confined 2DHG with HH–LH mixing and nonparabolic subband dispersions. Therefore, the sub band energies and splittings can differ from the Bir–Pikus estimate. The obtained splitting is only to demonstrate that the simulated strain is sufficiently large to modify the valence-band structure on the meV scale.

In addition to the biaxial strain component, significant shear strain develops near the gate edges, as shown in Fig. \ref{fig:Simulations}(b). To quantify the magnitude of the shear strain, we define

\begin{eqnarray}
\epsilon_{rot}=\sqrt{\epsilon_{xz}^{2}+\epsilon_{yz}^{2}},
\end{eqnarray}

which reaches values comparable to the biaxial strain invariant in localized regions near the gate perimeter. Within the Bir-Pikus formalism, such shear strain contributes off-diagonal coupling terms and can promote HH/LH hybridization.

\begin{acknowledgments}
We acknowledge the support of the HFML-FELIX, member of the European Magnetic Field Laboratory (EMFL).

The authors acknowledge support from the Netherlands Organization of Scientific Research (NWO) under VICI grant VI.C.222.083.

We would also like to thank Melissa Goodwin for making the STEM image.
\end{acknowledgments}

\bibliography{cleaned_references}

\end{document}